\pdfoutput=1

\documentclass{moriond}
\usepackage{amsmath}
\usepackage{subfig}
\usepackage{hyperref}

\def\be{\begin{equation}}
\def\ee{\end{equation}}
\def\bea{\begin{eqnarray}}
\def\eea{\end{eqnarray}}

\newcommand{\Photo}{
\includegraphics[height=35mm]{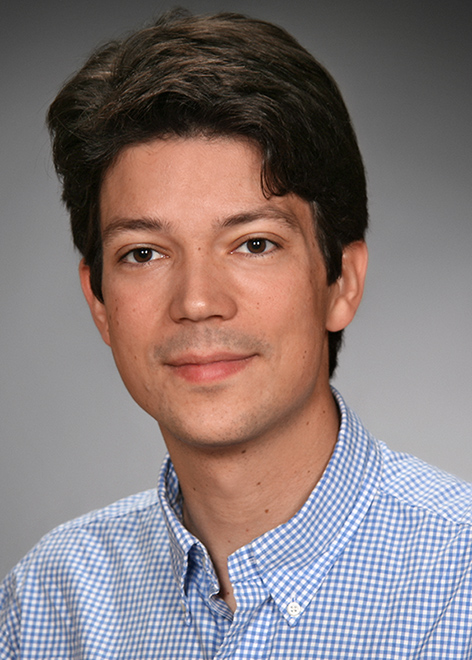}}

\begin{document}
\vspace*{4cm}
%\title{MULTILOOP INTEGRALS MADE SIMPLE: \\APPLICATIONS TO QCD PROCESSES}
\title{Multiloop integrals made simple: applications to QCD processes}

\author{ JOHANNES M. HENN}

\address{Institute for Advanced Study, \\
      Einstein Drive, Princeton, NJ 08540, USA\\
{E-Mail: {\tt jmhenn@ias.edu}}      
      \vspace*{0.3cm}
      }

\maketitle\abstracts{
I will present a new method for thinking about and for computing loop
integrals based on differential equations. All required information is
obtained by algebraic means and is encoded in a small set of simple
quantities that I will describe. I will present various applications,
including results for all planar master integrals  that are needed for the
computation of NNLO QCD corrections to the production of two off-shell
vector bosons in hadron collisions.}

\section{Introduction}

The theoretical description of particle collisions at the LHC
relies on precise calculations of the underlying microscopic scattering processes. 
As D.~Kosower has reviewed \cite{proceedingsKosower}, computing scattering amplitudes and cross sections at NLO (next-to-leading order) is by now completely standard and largely automated. There are various ingredients that make this possible, in particular efficient techniques for generating and organizing integrands, knowledge of one-loop Feynman integrals, and subtraction methods. Several talks at this conference reported on results that were obtained using automated NLO programs.

It is clear from the experimental results presented at this conference \cite{proceedingsCampana} that we have entered an era of precision measurements. Increasing experimental accuracy needs to be matched by theory predictions, and as a consequence, going beyond the NLO level is often necessary. 
 
For select processes, a full NNLO analysis is already available, see e.g. M.~Grazzini's talk \cite{proceedingsGrazzini}.
Often, however, the theoretical bottleneck is missing analytic results for the virtual 
loop integrals. In this talk I will report on a breakthrough in understanding and computing loop integrals that will help to close this gap.

This talk is based on ref. \cite{Henn:2013pwa}, where the main ideas where presented, and several papers with applications \cite{Henn:2013tua,Henn:2013woa,Henn:2013nsa,Henn:2014lfa}.
These advances are largely based on a better understanding of Feynman integrands, before integration, and the connection to the integrated functions. An important aspect 
of our analysis is that the class of special functions required for each type of Feynman integral is readily identified, and the analytic answer is then computed in terms of those functions.
The calculation itself is algebraic, and the answer can be presented in a pleasingly simple form.
In fact, the integrals are essentially specified by their singularity structure, which as a consequence is made
very transparent. Physical limits, such as Regge or threshold limits are very easy to discuss. 
Also, the method is especially useful for integrals that depend on several scales, which are hard to tackle in
traditional approaches. As we will see, it has been applied successfully to non-trivial integrals that can be massless, massive, planar or non-planar.

The method uses differential equations, whose application to Feynman integrals has a long and successful history, see \cite{Kotikov:1991pm,Remiddi:1997ny,Gehrmann:2000zt,proceedingsGehrmann}. This work is a refinement of this method.

\section{Sample applications}

\begin{figure}[tbp]
  \centering
  \subfloat[]{\includegraphics[width=0.22\textwidth]{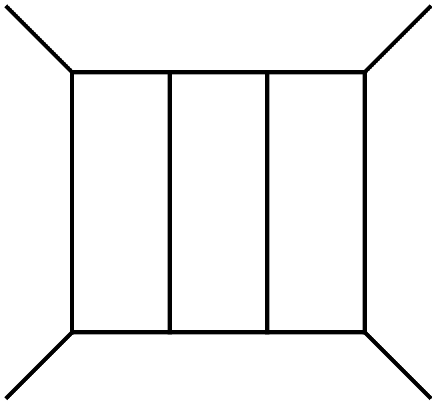}}
  \hspace{2cm}
  \subfloat[]{\includegraphics[width=0.2\textwidth]{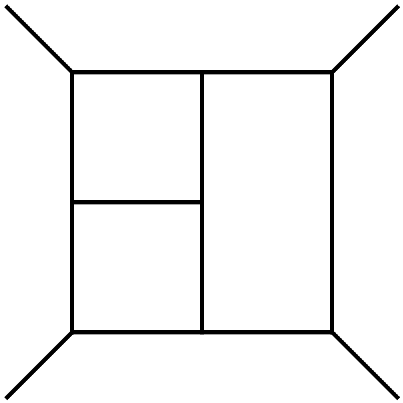}}
  \hspace{2cm}
  \subfloat[]{\includegraphics[width=0.3\textwidth]{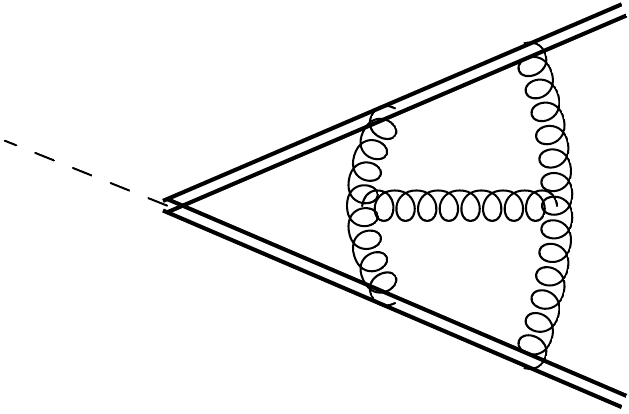}}  
  \caption{Examples of two-scale integral families computed using the new method. 
  Figures (a) and (b) show three-loop massless $2 \to 2$ integrals that depend on a dimensionless variable $x=t/s$. Fig. (c) shows Wilson line integrals in HQET that depend on the cusp angle $\phi$.}
  \label{figure_examples1}
\end{figure}

Before continuing to talk about the method itself, let me show you some examples of 
loop integrals that have been computed using it.
I hope that this will convince you that the method can be used for nontrivial calculations.
Each of the cases has its own interesting physics motivation, 
which unfortunately I can only hint at here due to time limitations.

The examples are naturally organized by the number of scales involved. Let us start with two-scale problems,
i.e. problems that depend on one dimensionless variable.
In a first nontrivial example, three-loop planar massless four-point integrals
were evaluated \cite{Henn:2013tua}.
These integrals depend on two scales, the Mandelstam variables $s$ and $t$.
See Fig.~\ref{figure_examples1}(a) and \ref{figure_examples1}(b). 
It should be noted that these figures represent whole families of integrals, 
in the sense that a basis for all integrals of this type
was computed.
For example, the basis for the family shown in Fig.~\ref{figure_examples1}(a) consists of $26$ integrals that include integrals with missing propagators, or with numerator factors inserted.
The knowledge of this basis allows to write down an analytic formula for any integral
of this family. 

Further results \cite{Henn:2013nsa} showed that the method applies equally to non-planar integrals, and work towards computing all non-planar three-loop integrals is in progress. Possible applications include the study of non-planar scattering amplitudes in Yang-Mills and (super)gravity theories.

As a second example, in work in progress \cite{GHKM} the planar three-loop cusp integrals in heavy quark effective theory (HQET) were computed.
See Fig.~\ref{figure_examples1}(c) for a sample integral. Here the physical motivation is the study of the structure infrared divergences of massive scattering amplitudes in QCD.

\begin{figure}[tbp]
  \centering
  \subfloat[]{\includegraphics[width=0.3\textwidth]{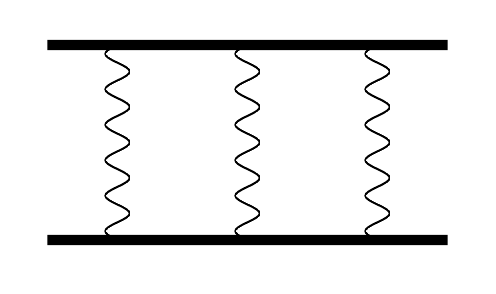}}
  \hspace{1cm}
  \subfloat[]{\includegraphics[width=0.3\textwidth]{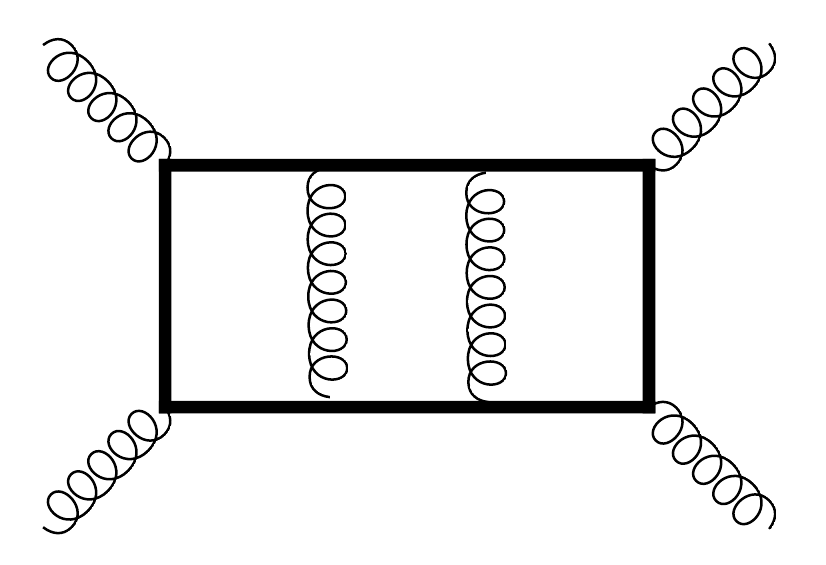}}
  \hspace{1cm}
  \subfloat[]{\includegraphics[width=0.2\textwidth]{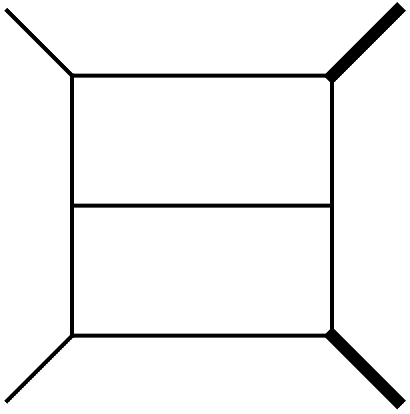}}
  \caption{Examples of multi-scale integral families computed using the new method. 
  Fat lines correspond to massive particles (or off-shell vector bosons).
   (a) Integrals for Bhabha scattering.
     (b) Massive integrals depending on $3$ scales.
   (c) Planar integrals for vector boson production.}
  \label{figure_examples2}
\end{figure}

Let us now move on to multi-scale problems. 
%It turns out that here
This is probably where the advantages of the new method
can be best seen. 
%One reason is simply that such problems were previously considered to be 
%very complicated. Another reason is that the appropriate class of required functions
%is readily identified within our approach.

A first example is a family of integrals appearing in Bhabha scattering \cite{Henn:2013woa}, which depend on $s,t$, and the mass $m$. 
See Fig.~\ref{figure_examples2}(a). 
Staying with $3$ scales, but moving up in the loop order, to three loops, we have a class of integrals that appear in light-by-light scattering \cite{Caron-Huot:2014lda}, cf. Fig.~\ref{figure_examples2}(b). 
Finally, coming back to integrals directly relevant for current 
LHC physics, in \cite{Henn:2014lfa} all planar
integral families for the description of the production of two off-shell vector bosons 
in hadron collisions at NNLO were computed. This is a four-scale problem,
depending on $s,t,M_a^2, M_b^2$. A sample class of integrals is shown in 
Fig.~\ref{figure_examples2}(c).  
Our results generalize the equal mass case of ref. \cite{Gehrmann:2013cxs}.

\section{Main features of the method}

Let me describe the main features of the method. 
A given scattering process is described by a set of master, or basis integrals $\vec{f}$. 
The first step consists in deriving differential equations (DE) in the kinematic variables. The crucial new idea is to choose a convenient basis, in which the differential equations take a simple, standard form. 
Criteria for finding such a basis that can be systematically applied were given in ref. \cite{Henn:2013pwa}, and examples given in ref. \cite{Henn:2013tua}.
As we will see, once a good basis is chosen, the solution for $\vec{f}$ can be immediately read off from the DE.

For example, for integrals depending on one dimensionless variable $x$ (and on the dimension $D=4-2 \epsilon$), the standard form proposed in \cite{Henn:2013pwa} is
\begin{equation}\label{exampelDE}
\partial_x \vec{f}(x;\epsilon) = \epsilon \sum_k \frac{A_k}{x-x_k } \vec{f}(x;\epsilon) \,.
\end{equation}
{}From this we see that the data defining $\vec{f}$ is elegantly described by a set of {\it{ letters}} ${\alpha = \{ x-x_{k}\}}$, 
related to the singularities $x_{k}$, and a corresponding set of constant $N \times N$ matrices $A_{k}$. Here $N$ is the size of the basis. 

This data specifies the class of functions that the answer can be expressed in.
In fact, expanding the solution for $\vec{f} = \sum_k \epsilon^k \vec{f}^{(k)}$ to any order in $\epsilon$ amounts to linear algebra,
and the term at order $\epsilon^{k}$ is given in terms of multiple polylogarithms of uniform weight (colloquially referred to as `transcendentality') $k$. These iterated integrals are generalizations of logarithms and dilogarithms and 
have very nice mathematical properties.
Thanks to the uniform weight property results are much more compact compared to other 
approaches.

Moreover, the analytic behavior of the answer is very transparent from the DE \cite{Henn:2013nsa}.
For example, the asymptotic behavior can be easily read off, e.g. $\vec{f}(x;\epsilon) \sim (x-x_{k})^{\epsilon A_k } \vec{f}_{0}(\epsilon)$, which is helpful e.g. when fixing the boundary conditions.

The method has a natural extension to the multi-variable case, which we discuss next,
using the $q \bar{q}\longrightarrow V_1 V_2$ integrals mentioned earlier as an example. 

\section{Example: vector boson production at NNLO}

The planar master integrals can be organized into three classes, one of which is shown in 
Fig.~\ref{figure_examples2}(c). There are of the order of $N \approx 30$ integrals for each case.
The integrals depend on three dimensionless variables $x,y,z$ related to $s,t,M_a^2, M_b^2$ via
\begin{align}
s/M_{a}^2 = (1+x)(1+x y)\,,\quad t/M_{a}^2 = -x z \,,\quad M_{b}^2/M_{a}^2 = x^2 y \,.
\end{align} 
The DE w.r.t. those variables can be compactly written in
differential form,
\begin{equation}\label{DEVV}
d\, \vec{f}(x,y,z;\epsilon) = \epsilon \, d\,\tilde{A}(x,y,z) \vec{f}(x,y,z;\epsilon) \,, \qquad \tilde{A} = \sum_{i=1}^{15} \tilde{A}_{\alpha_{i}} \log(\alpha_i ) \,.
\end{equation}
Here the basis choice for $\vec{f}$ was straightforwardly made using the criteria of ref. \cite{Henn:2013pwa}.

The {\it {alphabet}} appearing in the DE is given by a set
of rational functions of $x,y,z$. Specifically, we find 
\begin{equation}
\label{alphabet}
\begin{aligned}
 \alpha =& \{ x, y, z, 1 + x, 1 - y, 1 - z,  1 + x y, z-y,
1 + y(1 + x) - z,  x y + z,  \\
&
 1 + x(1 +  y - z), 1 + x z,  1+y-z, z + x (z - y) + x y z ,   z-y + y z + x y z \}.
\end{aligned}
\end{equation}
This alphabet tells us which class of iterated integrals appear in the answer.
It reflects the rich singularity structure of this many-scale scattering process.
As before, the answer is written in terms of multiple polylogarithms, at any order in $\epsilon$. Dedicated computer codes for their efficient numeric evaluation are available \cite{Bauer:2000cp,Vollinga:2004sn}.
For convenience, the result up to weight four, which is the order needed for NNLO calculations, is provided in electronic form in \cite{Henn:2014lfa}.
Finally we wish to mention that, in addition to making the singularity structure of these functions completely manifest, eq. (\ref{DEVV}) also makes it trivial to obtain series expansions in kinematical limits.  

\section{Conclusion}

In this talk I have presented results for a number of non-trivial families of Feynman integrals that are described by iterated integrals. The method used to compute them uses only minimal data to specify the analytic answer, namely
\begin{itemize}
\item[(a)] an {\it{alphabet}} $\alpha$ for iterated integrals, see eq. (\ref{alphabet}). 
\item[(b)] rules for forming words in this alphabet. These rules are provided by a
set of constant matrices $\tilde{A}_{\alpha_{i}}$, see eq. (\ref{DEVV}).
They determine which linear combinations of integrals precisely constitute the answer,
at any order in the $\epsilon$ expansion.
\end{itemize}
This reminds me of a short story by Borges\cite{Borges}, {\it{La biblioteca de Babel}}, where the author imagines an infinite library, whose books are composed from infinite random sequences of letters. Similarly, for Feynman integrals the $\epsilon$ expansion can be driven to any desired order, generating longer and longer expressions, which are words in the alphabet $\alpha$. However, the latter derive from the simple data described above. In particular, the constant matrices provide the `grammar' for forming words, as opposed to the random generator in Borges' imaginary library.

Returning from magical libraries to real (or digital) ones,
as I have described, this method of analyzing and computing loop integrals has already been applied successfully to cases relevant to current LHC physics. 
I think as an outlook one can envisage a library for NNLO Feynman integrals
for phenomenology. In this spirit, do not hesitate to contact me if there are new 
integrals that you are interested in.

Although not the main focus at this conference, I wish to mention that there are 
interesting connections to several fields of mathematics and mathematical physics, 
and I think this is worth exploring further. Also, there are interesting open question for integrals containing elliptic functions.

Finally, I cannot resist making an advertisement: If you are curious about the method I have described, I invite you to look at sections 3.8 and 3.9 of 
the recent volume of lecture notes \cite{Henn:2014yza}, which include a pedagogical introduction, 
using the gluon fusion process as an example.

\section*{Acknowledgments}
J.M.H. is supported in part by DOE grant DE-SC0009988, and by  
a Marvin L. Goldberger membership at the Institute for Advanced Study.


\newpage 

\section*{References}

\begin{thebibliography}{99}

\bibitem{proceedingsKosower}
David Kosower,
\newblock {these proceedings}.

\bibitem{proceedingsCampana}
Pierluigi Campana,
\newblock {these proceedings}.

\bibitem{proceedingsGrazzini}
Massimiliano Grazzini,
\newblock {these proceedings}.

\bibitem{Henn:2013pwa}
Johannes~M. Henn,
\newblock {\em Multiloop integrals in dimensional regularization made simple},
\newblock {\em Phys.Rev.Lett.}, 110:251601 (2013).

\bibitem{Henn:2013tua}
Johannes~M. Henn, Alexander~V. Smirnov, and Vladimir~A. Smirnov,
\newblock {\em  Analytic results for planar three-loop four-point integrals from a
  Knizhnik-Zamolodchikov equation},
\newblock {\em JHEP}, 1307:128 (2013).

\bibitem{Henn:2013woa}
Johannes~M. Henn and Vladimir~A. Smirnov,
\newblock {\em  Analytic results for two-loop master integrals for Bhabha scattering
  I},
\newblock {\em JHEP}, 1311:041 (2013).

\bibitem{Henn:2013nsa}
Johannes~M. Henn, Alexander~V. Smirnov, and Vladimir~A. Smirnov,
\newblock {\em  Evaluating single-scale and/or non-planar diagrams by differential
  equations},
\newblock {\em JHEP}, 1403:088 (2014).

\bibitem{Henn:2014lfa}
Johannes~M. Henn, Kirill Melnikov, and Vladimir~A. Smirnov,
\newblock {\em  Two-loop planar master integrals for the production of off-shell
  vector bosons in hadron collisions},
\newblock {arXiv:1402.7078}.

\bibitem{Kotikov:1991pm}
A.~V. Kotikov,
\newblock {\em  Differential equation method: The Calculation of N point Feynman
  diagrams},
\newblock {\em Phys. Lett.}, B267:123--127 (1991).

\bibitem{Remiddi:1997ny}
Ettore Remiddi,
\newblock {\em  Differential equations for Feynman graph amplitudes},
\newblock {\em Nuovo Cim.}, A110:1435--1452 (1997).

\bibitem{Gehrmann:2000zt}
T.~Gehrmann and E.~Remiddi,
\newblock {\em  Two-Loop Master Integrals for $\gamma^* \to 3$ Jets: The planar
  topologies},
\newblock {\em Nucl. Phys.}, B601:248--286 (2001).

\bibitem{proceedingsGehrmann}
Thomas Gehrmann,
\newblock {these proceedings}.

\bibitem{GHKM}
Andrey Grozin, Johannes~M. Henn, Gregory Korchemsky, and Peter Marquard,
\newblock {\em to appear} (2014).

\bibitem{Caron-Huot:2014lda}
Simon Caron-Huot and Johannes~M. Henn,
\newblock {\em  Iterative structure of finite loop integrals},
\newblock  arXiv:1404.2922.

\bibitem{Gehrmann:2013cxs}
Thomas Gehrmann, Lorenzo Tancredi, and Erich Weihs,
\newblock {\em Two-loop master integrals for $q \bar{q} \to VV$: the planar
  topologies},
\newblock {\em JHEP}, 1308:070 (2013).

\bibitem{Bauer:2000cp}
Christian~W. Bauer, Alexander Frink, and Richard Kreckel,
\newblock {\em Introduction to the GiNaC framework for symbolic computation within
  the C++ programming language},
\newblock unpublished, (2000).

\bibitem{Vollinga:2004sn}
Jens Vollinga and Stefan Weinzierl,
\newblock {\em Numerical evaluation of multiple polylogarithms},
\newblock {\em Comput.Phys.Commun.}, 167:177 (2005).

\bibitem{Borges}
Jorge~Luis Borges,
\newblock {\em El Jard\'in de senderos que se bifurcan},
\newblock {\em Editorial Sur} (1941).

\bibitem{Henn:2014yza}
Johannes~M. Henn and Jan~C. Plefka,
\newblock {\em Scattering Amplitudes in Gauge Theories},
\newblock {\em Lect.Notes Phys.}, 883 (2014).

\end{thebibliography}
\end{document}